\documentclass[reprint,prl,aps,floatfix,unsortedaddress]{revtex4-1}
\usepackage{graphicx}
\usepackage{amsmath,amsfonts,mathcomp}
\usepackage[modulo]{lineno}
\usepackage{siunitx}
\usepackage{xcolor,colortbl,tabularx,multirow}
\usepackage{listings}
\usepackage[english]{babel}
\usepackage{array}
\newcolumntype{L}[1]{>{\raggedright\let\newline\\\arraybackslash\hspace{0pt}}m{#1}}
\newcolumntype{C}[1]{>{\centering\let\newline\\\arraybackslash\hspace{0pt}}m{#1}}
\newcolumntype{R}[1]{>{\raggedleft\let\newline\\\arraybackslash\hspace{0pt}}m{#1}}

\begin{document}
\title{The small impact of various partial charge distributions in ground and excited state on the computational Stokes shift of 1-methyl-6-oxyquinolinium betaine in \textcolor{black}{diverse water models}}
\author{Esther Heid, Sophia Harringer and Christian Schr\"oder}
\email{christian.schroeder@univie.ac.at}
\affiliation{University of Vienna, Faculty of Chemistry, Department of Computational Biological Chemistry, 
W\"ahringerstra{\ss}e 19, A-1090 Vienna, Austria}

\begin{abstract}
The
influence of the partial charge distribution obtained from quantum mechanics  of the solute 1-methyl-6-oxyquinolinium betaine in the ground- and first excited state 
on the time-dependent Stokes shift  is studied via molecular dynamics computer simulation. Furthermore,
the effect of the employed solvent model -- here the non-polarizable SPC, \textcolor{black}{TIP4P and TIP4P/2005} and the polarizable SWM4 water model -- on the solvation dynamics of the system is investigated.
The use of different functionals and calculation methods influences the partial charge distribution and the magnitude of the dipole moment of the solute, but not the orientation of the dipole moment. Simulations based
on the calculated charge distributions show nearly the same relaxation behavior. Approximating the whole solute molecule by a dipole results in the same relaxation behavior,
but lower solvation energies, indicating that the time scale of the Stokes shift does not depend on peculiarities of the solute. However, \textcolor{black}{the SPC and TIP4P water models} show too 
fast dynamics which can be ascribed to a too large diffusion coefficient and 
too low viscosity. The calculated diffusion coefficient and viscosity for the SWM4 and TIP4P/2005 model coincide well with experimental values and the corresponding relaxation behavior is comparable to experimental values.
Furthermore we found that for a quantitative description of the Stokes shift of the applied system at least two solvation shells around the solute have to be taken into account.
\end{abstract}
\maketitle

\section{Introduction}
Solvation dynamics spectroscopy monitors the solvent response to an electronic excitation of a chromophore and provides useful information on the dynamics of the interactions between the solute and 
its surrounding solvent molecules. The solvent relaxation behavior and its timescale are important for the rate of chemical reactions in that solvent since a retarded solvent response to the electronic 
rearrangement of solute molecules passing the transition state may result in free energy barriers reducing the reaction rate~\cite{mar94a}.
Moreover, solvation plays an important role in biomolecular function~\cite{bag00b}. Consequently, chromophores attached to proteins, to DNA~\cite{ber09a} or to trehalose~\cite{ern14a} allow for experimental studies 
on water dynamics near biomolecular surfaces.

After electronic excitation of the solute the fluorescence spectrum changes in time as the solvent molecules reorganize. Often, the time evolution of the maximum of the fluorescence band $\nu(t)$
is reported in terms of a normalized spectral relaxation function $S(t)$, \textit{i.e.} the Stokes shift,
\begin{equation}
S(t) = \frac{\nu(t) - \nu(\infty)}{\nu(0) - \nu(\infty)}
\label{EQU:S}
\end{equation}
which shows bimodal behavior~\cite{mar93a,mar94a,mar12b,pet04a, sam02a, sam03a} in many solvents ranging from femtosecond dynamics in water~\cite{mar94a,bag00b,ern11a} to nanoseconds in ionic liquids 
\cite{mar03a,mar07a,bis13a}.

Although solvation dynamics is about studying solvents, the chromophore used to probe the solvation dynamics seems to have an influence on $S(t)$ as well. 
For example, Ernsting and coworkers measured the solvent response of 1-methyl-6-oxyquinolinium betaine (1MQ), Coumarin 153 and 343, as well as 5 other solutes in water, methanol and benzonitril
\cite{ern11a} and reported that in methanol the average relaxation time 
\begin{equation}
 \langle \tau \rangle = \int\limits_0^\infty S(t) dt
\end{equation}
of $S(t)$ increases from \SI{2.7}{\pico\second} to \SI{6.4}{\pico\second} when going from 1MQ to 4-aminophthalimide. However, the solvation
dynamics in water is much faster ($\langle \tau \rangle \approx$~\SI{0.45}{\pico\second}) and shows no dependence on the nature of the chromophore.
Horng \textit{et al.} found  that chromophores with strong hydrogen bonding networks show significantly slower solvation dynamics in 1-propanol 
as the majority of their investigated chromophores ~\cite{mar95a}.

\bigskip
However, to investigate the relaxation dynamics of various systems more thoroughly, it is necessary to combine experimental research with computer simulations.
Ultrafast components below the limit of experimental resolution cannot be measured reliably for some solvents like water, and it was only through simulation
that the inertial component of the solvation response could be examined in many common liquids~\cite{mar98a}. Computer simulation gives also information about the translation of solute 
and solvent molecules and therefore about the individual contributions to the overall function. After excitation, different processes on various time scales start to happen:
The electrons of the solvent molecules adjust to the new partial charge distribution of the solute, which is too fast to be measured in experiment. Also, the intramolecular bonds in the solvent can 
be distorted slightly on a vibrational time scale. The largest contribution comes from reorientation of the solvent molecules through rotation and translation on a picosecond timescale, 
or non-diffusive libration which is much faster ~\cite{mar93a}.
This gives rise to the need of computer simulation of dynamic solvation to gain deeper understanding of the processes taking place after solute excitation.

\subsection{Computer simulation of 1MQ in water}
The current computational work is a pilot study concerning the solvation dynamics of oxyquinolinium betaine and serves as a starting point for a series of subsequent
molecular dynamics (MD) simulations of various oxyquinolones in various solvents. 1MQ has been used for studying solvation dynamics~\cite{ern05a,ern11a,ern14a} since it is rather small and soluble in water.
It has no net charge and is rigid, so that it will not interfere with the vibrational modes of water~\cite{seb11a}. In contrast to the standard chromophore coumarin C153, 1MQ reduces 
its dipole moment upon laser excitation. Thus, general conclusions on solvation dynamics can be tested for chromophores weakening their local electric field when going from ground 
$\mathbb{S}_0$ to excited state $\mathbb{S}_1$.
Furthermore, it can be attached to biomolecules~\cite{ern14a} and also easily modified to introduce various moieties at various positions changing the shape, 
volume and hydrogen bonding capabilities of the solute. The impact of these modifications will be topic of subsequent publications.

\bigskip
Computational studies on the solvation dynamics of 1MQ in water were performed by Sebastiani and coworkers~\cite{seb11a,seb13a}, who 
followed  the time-dependent Stokes shift $S(t)$ of 1MQ in up to 130 water molecules by ab initio MD simulations and found very good agreement to experimental data.
However, due to the enormous computational effort of ab initio MD, only few independent simulations restricted to a few picoseconds can be run.
Adding larger moieties to the oxyquinolinium betaine chromophore makes the ab initio MD simulations more tedious or even unfeasible. Also replacing water by more viscous 
solvents like ionic liquids renders the calculations impossible as this necessitates longer trajectories since $\langle \tau \rangle$ may reach the nanosecond timescale. 
This is where classical non-equilibrium MD simulations become important as they offer the possibility to produce trajectories for several nanoseconds, can easily deal with large solutes 
and large numbers of solvent molecules. Our parameter-free Voronoi analysis~\cite{ste09d} shows that on average 46 water molecules can be found in the first solvation shell around 1MQ. 
The second solvation shell already contains 127 water molecules. In other words, the above mentioned ab initio simulations do not contain a full second hydration shell. 
Since dielectric effects extend beyond the first hydration shell~\cite{seb13a}, the simulation of larger boxes seems inevitable to have at least a few water molecules behave like bulk water.
Moreover, the number of independent simulations can be increased to 1000 when using MD simulations as presented in this work. 
Ab initio and classical MD simulations can therefore be used to gather complementary information: Ab initio MD provides very accurate results for small, non-viscous systems through 
precise treatment of the solute, whereas classical MD simulations can be applied for large or highly viscous systems albeit the drawback of using some simplistic assumptions on the solute. 
As already mentioned above, this work serves as the starting point of the simulation of alterated oxyquinolones, some of them very large, in both high- and low viscous solvents, which
makes the use of classical MD instead of ab initio calculations inevitable. 

\subsection{Computation of the Stokes shift from non-equilibrium MD simulations}
Classical non-equilibrium MD simulations rely on the assumption that the interaction energy between the chromophore and the solvent is of electrostatic nature,
\textit{i.e.} the relative Stokes shift can be computed by
\begin{equation}
S(t) = \frac{\Delta U(t) - \Delta U(\infty)}{\Delta U(0) -\Delta U(\infty)}.
\end{equation}
using change of the Coulomb energy 
\begin{equation}
\Delta U(t) = \frac{1}{4 \pi \epsilon_0} \sum\limits_{j\gamma} \sum\limits_{i\beta} \frac{\Delta q_{j\gamma} \cdot q_{i\beta}}{r_{j\gamma i\beta}(t)}
\label{EQU:DeltaU}
\end{equation}
between the chromophore atoms $\gamma$ of solute molecule $j$ (here, only one solute molecule is present) 
and the solvent atoms $\beta$ of molecule $i$ at distance $r_{j\gamma i\beta}$ when changing the partial charge distribution
 from ground $\mathbb{S}_0$ to excited state $\mathbb{S}_1$ by $\Delta q_{j\gamma}$. 
Another approach is  to approximate $\Delta U(t)$ via the change in dipole moment $\Delta \vec \mu_j$ of the solute and the reaction field $\vec E_j^{\mathrm{RF}}(t)$,
so that the interaction energy becomes
\begin{eqnarray}
\Delta U(t)  &=& - \sum_j \Delta \vec \mu \cdot \vec E_j^{\mathrm{RF}}(t) \\
&=& \frac{1}{4 \pi \epsilon_0} \sum_j \Delta \vec \mu_j \sum\limits_{i\beta} \frac{q_{i\beta}\cdot  \vec r_{j i\beta}(t)}{r_{j i\beta}^3(t)}
\label{EQU:DeltaUq}
\end{eqnarray}
where $\vec r_{j i\beta}(t)$ is the vector from the center of mass of the solute molecule $j$ to atom $\beta$ of solvent molecule $i$. 
In this study we will investigate whether this approximation holds true for the 1MQ - water system, \textit{i.e.} the oxyquinolinium betaine molecule behaves
similar to a dipole in a (nearly) spherical cavity surrounded by water molecules.

\bigskip
Inherently in classical MD simulations, the partial charges do not change during 
the simulation. Consequently, the non-equilibrium simulations assume the chromophore to be in the excited state until the Stokes shift relaxation has taken place.
Furthermore, and even more important, neutral solvent molecules in classical MD simulations have also fixed partial charges $q_{i\beta}$. 
In polarizable MD
simulations by means of Drude oscillators, however, the non-hydrogen atoms of the solvent can be made polarizable, so that the induced dipoles may react ultrafast to the changing local 
electric field exerted by the solute. We already showed for the chromophore coumarin C153 in ionic liquids ~\cite{ste13a} that the cross-correlation between 
the induced and permanent contributions play an important role for the Stokes shift. Furthermore, polarizable force field models better reproduce experimental 
physico-chemical properties of the solvent, \textit{e.g.} for ionic liquids~\cite{bor09a,sch12a} or water~\cite{hes02a,ker05c}. In the present work we 
demonstrate better agreement to the experimental Stokes shift of 1MQ in water when using the polarizable water SWM4 model~\cite{rou03b} compared to the 
non-polarizable SPC~\cite{her81a} and \textcolor{black}{TIP4P~\cite{kle83a} water models}.

\subsection{Partial charge distribution of the solute}
One prerequisite of the computation of $S(t)$ by MD simulations is the partial charge distribution in ground and excited state of the chromophore 
which have to be determined quantum mechanically. In this work we study the impact of the various partial charge distribution 
$\Delta q_{j\gamma}$ gained from various functionals with and without a polarizable continuum model to take the solvent implicitly into account.
Furthermore, we also test two different procedures to assign partial charges to particular atoms, namely CHelpG~\cite{wib90a} and Hirshfeld~\cite{hir77a}.

\section{Methods}
The partial charge distribution of 1MQ was calculated using DFT/hybrid DFT for the ground state and TD-DFT for excited states in Gaussian 09~\cite{gau09a}, where we chose
the B3LYP DFT functional~\cite{bec88a,lee88a}, the PBE0 hybrid DFT functional~\cite{per96a} and the $\omega$B97xD hybrid DFT functional~\cite{cha08a}. 
The PBE0 functional has been shown to yield accurate excitation energies, unlike most of other hybrid DFT functionals, for organic dyes like 1MQ~\cite{ada09a}.
\textcolor{black}{Despite the known problems of TD-DFT for the calculation of charge-transfer states~\cite{gor04a,hut04a}, Sebastiani and coworkers calculated the excited state of 1MQ comparing the TD-DFT method to
CIS, ROKS and EOM-CCSD and found surprisingly good agreement of the TD-DFT approach with the more sophisticated methods~\cite{seb11a}, so that we will use the computationally cheaper TD-DFT.}
An aug-cc-pVTZ basis set was employed for all methods. 
All calculations were done in vacuum as well as in implicit solvent using the polarizable continuum model (PCM) of water~\cite{tom05b}. 
\begin{figure}[t]
  \includegraphics [width=6cm]{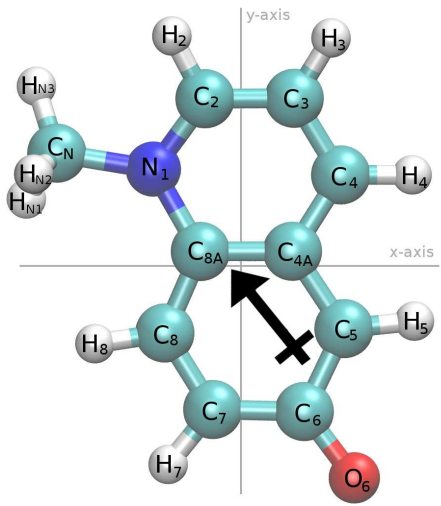}
  \caption{Dipole moment of the ground state of 1-methyl-6-oxyquinolinium betaine calculated using CHelpG B3LYP in vacuum. Origin of the axes is located at the center of mass.
  \label{FIG:dipole}}
\end{figure}

\bigskip
The structure of 1MQ depicted in Fig.~\ref{FIG:dipole} was optimized on the B3LYP 6-311G++(2d,2p) level of theory, a subsequent frequency calculation was done to verify the geometry as a true minimum. 
The respective partial charge distributions were evaluated using either the CHelpG~\cite{wib90a} or the Hirshfeld method~\cite{hir77a}. The CHelpG method calculates the partial charges from the molecular
electrostatic potential using a grid-based method. It is therefore independent from molecular orientation (unlike the former CHelp method), but works only for rather small molecules. In contrast, the Hirshfeld method dissects the molecule
into atomic fragments and assigns partial charges according to free-atom densities.

\bigskip
For the computation of the time-dependent Stokes shift relaxation function $S(t)$ we performed 1000 independent non-equilibrium MD simulations in CHARMM~\cite{kar09a} 
for each of the above mentioned partial charge distributions.
The non-polarizable force field of 1MQ including intra- and intermolecular potentials was obtained from PARAMCHEM~\cite{van12a,van12b} which is based on the CHARMM General Force Field (CGenFF)~\cite{van10a}, where we replaced the partial charges with our charge 
distributions. The respective force field parameters of 1MQ are given in the supplementary material~\cite{esi}. Polarizability of the solute was not taken into account, since atomic polarizabilities for the excited state are unknown.

\bigskip
Molecular dynamics models of the non-polarizable SPC water~\cite{her81a}, \textcolor{black}{TIP4P~\cite{kle83a}, TIP4P/2005~\cite{veg05a}} and the polarizable SWM4-NDP water~\cite{rou03b} model were used for the solvent.
The initial configurations were generated by randomly packing one molecule of 1MQ and 1000 water molecules in a cubic box with a length of \SI{32}{\angstrom} using PACKMOL~\cite{mar09a}.
During a NpT equilibration for \SI{100}{\pico\second} at \SI{1}{bar} and \SI{300}{\kelvin} using an \textcolor{black}{Nos\'e}-Hoover thermostat~\cite{nos84a,hoo85a} the box length converged to \SI{31.05}{\angstrom} for SPC, \textcolor{black}{to \SI{31.23}
for TIP4P and TIP4P/2005} and 
to \SI{31.20}{\angstrom} for SWM4 water. 
A long NVT run at elevated temperatures was used to produce 1000 independent configurations of the system. 
These starting configuration replica were then used to simulate the electronic excitation of the solute molecule by the following protocol:
\begin{enumerate}
\item Equilibration (NVT ensemble) of the ground state for \textcolor{black}{\SI{100}{\pico\second} (SPC) or \SI{500}{\pico\second} (SWM4, TIP4P, TIP4P/2005)} at $T$=\SI{300}{\kelvin}. 
\item Instantaneous change of the partial charge distribution to the excited state without further change of other parameters like force constants or equilibrium geometry. 
\item NVT simulation for another \SI{50}{\pico\second} with a time step $\Delta t$ of \SI{1}{\femto\second} at $T$=\SI{300}{\kelvin}, where the coordinates were saved each femtosecond at the beginning
of the trajectory and then in intervals of 10, 100 and \SI{500}{\femto\second} respectively to save disk space.
\end{enumerate}

All simulations were carried out in cubic boxes with periodic boundary conditions. 
Energy calculation was done using the Particle Mesh Ewald method with grid size of nearly \SI{1}{\angstrom},
cubic splines of order 6, a $\kappa$ of \SI{0.41}{\angstrom}$^{-1}$ and a cut-off for non-bonded energy terms of \SI{11}{\angstrom}.
The resulting trajectories of step 3 were analyzed using a self-written Python program based on MDAnalysis~\cite{bec11a} to calculate the time-dependent Stokes shift.

\section{Results and discussion}
\subsection{Partial charge distributions}
\begin{table*}[t]
\caption{Partial charges $q_{j\gamma}$ in the ground state $\mathbb{S}0$ and their change $\Delta q_{j\gamma}$ upon excitation to $\mathbb{S}1$, 
of 1MQ based on the CHelpG and the Hirshfeld method with the functionals B3LYP, PBE0 and $\omega$B97xD. Atom numbering as depicted in Fig~\ref{FIG:dipole}. For coordinates see the supporting information~\cite{esi}.}
\centering
\footnotesize
\begin{tabular}{ p{0.78cm}  R{0.95cm}R{0.96cm}R{1.05cm}R{0.96cm}
                 p{0.3cm}  R{0.95cm}R{0.96cm}R{1.05cm}R{0.96cm}
                 p{0.3cm}  R{0.95cm}R{0.96cm}R{1.05cm}R{0.96cm}
                 p{0.3cm}  R{0.95cm}R{0.96cm}}
\hline\hline
 &\multicolumn{4}{c}{CHelpG \textbf{B3LYP}} 
   && \multicolumn{4}{c}{CHelpG \textbf{PBE0}} 
   && \multicolumn{4}{c}{CHelpG  \textbf{$\omega$B97xD}}
   && \multicolumn{2}{c}{Hirshfeld  \textbf{PBE0}}\\[0.1cm]
   &\multicolumn{2}{c}{vacuum} & \multicolumn{2}{c}{PCM}
   &&\multicolumn{2}{c}{vacuum} & \multicolumn{2}{c}{PCM}
   &&\multicolumn{2}{c}{vacuum} & \multicolumn{2}{c}{PCM}
   &&\multicolumn{2}{c}{vacuum} \\\hline
   &  $q_{j\gamma}$/e & $\Delta q_{j\gamma}$/e& $q_{j\gamma}$/e & $\Delta q_{j\gamma}$/e
   && $q_{j\gamma}$/e & $\Delta q_{j\gamma}$/e& $q_{j\gamma}$/e & $\Delta q_{j\gamma}$/e
   && $q_{j\gamma}$/e & $\Delta q_{j\gamma}$/e& $q_{j\gamma}$/e & $\Delta q_{j\gamma}$/e
   && $q_{j\gamma}$/e & $\Delta q_{j\gamma}$/e    \\\hline
\textbf{C$_\mathrm{N}$} &-0.0624& 0.0042&-0.0569&-0.0367&&-0.1402& 0.0065&-0.1329&-0.0324&&-0.1187&-0.0351&-0.1027&-0.0763&&-0.1141&-0.0088 \\
\textbf{H$_\mathrm{N1}$}& 0.0714&-0.0202& 0.0841&-0.0111&& 0.0929&-0.0197& 0.1068&-0.0121&& 0.0891&-0.0109& 0.0996& 0.0004&& 0.1267&-0.0084\\
\textbf{H$_\mathrm{N2}$}& 0.0714&-0.0202& 0.0841&-0.0111&& 0.0929&-0.0197& 0.1068&-0.0121&& 0.0891&-0.0109& 0.0996& 0.0004&& 0.1267&-0.0084\\
\textbf{H$_\mathrm{N3}$}& 0.0714&-0.0202& 0.0841&-0.0111&& 0.0929&-0.0197& 0.1068&-0.0121&& 0.0891&-0.0109& 0.0996& 0.0004&& 0.1267&-0.0084\\
\textbf{N$_1$}          & 0.0215& 0.1377& 0.0226& 0.1187&& 0.0421& 0.1368& 0.0363& 0.1190&& 0.0340& 0.1510& 0.0273& 0.1115&&-0.2662&-0.0378\\
\textbf{C$_2$}          &-0.0533&-0.1533& 0.0156&-0.1921&&-0.0600&-0.1609& 0.0269&-0.2155&&-0.0434&-0.1842& 0.0231&-0.2082&& 0.0273& 0.0041\\
\textbf{H$_2$}          & 0.1082& 0.0241& 0.1354& 0.0109&& 0.1163& 0.0266& 0.1420& 0.0154&& 0.1157& 0.0250& 0.1473& 0.0107&& 0.1306&-0.0032\\
\textbf{C$_3$}          &-0.1567& 0.1142&-0.1617& 0.0812&&-0.1729& 0.1162&-0.1842& 0.0874&&-0.1813& 0.1054&-0.1788& 0.0587&&-0.1083&-0.0093\\
\textbf{H$_3$}          & 0.1156&-0.0242& 0.1363&-0.0257&& 0.1273&-0.0242& 0.1501&-0.0265&& 0.1314&-0.0240& 0.1511&-0.0245&& 0.1191&-0.0054\\
\textbf{C$_4$}          &-0.0845&-0.1609&-0.0832&-0.1600&&-0.0851&-0.1703&-0.0742&-0.1794&&-0.0834&-0.1666&-0.0863&-0.1541&&-0.0733&-0.0503\\
\textbf{H$_4$}          & 0.1093& 0.0015& 0.1281&-0.0118&& 0.1167& 0.0047& 0.1379&-0.0095&& 0.1218&-0.0036& 0.1433&-0.0156&& 0.1216&-0.0188\\
\textbf{C$_\mathrm{4A}$}& 0.1639& 0.0945& 0.1886& 0.0365&& 0.1427& 0.1026& 0.1592& 0.0513&& 0.1568& 0.0551& 0.1894&-0.0023&&-0.0144& 0.0184\\
\textbf{C$_5$}          &-0.5055& 0.2204&-0.5613& 0.2544&&-0.5001& 0.2156&-0.5421& 0.2317&&-0.5263& 0.2564&-0.5888& 0.2551&&-0.1443& 0.1017\\
\textbf{H$_5$}          & 0.1647&-0.0444& 0.1638&-0.0250&& 0.1704&-0.0412& 0.1689&-0.0200&& 0.1700&-0.0424& 0.1746&-0.0222&& 0.1050& 0.0183\\
\textbf{C$_6$}          & 0.6685&-0.0045& 0.7194&-0.0213&& 0.6432&-0.0002& 0.6782&-0.0034&& 0.6697&-0.0194& 0.7151&-0.0171&& 0.1058& 0.0174\\
\textbf{O$_6$}          &-0.6706& 0.0994&-0.8357& 0.1670&&-0.6563& 0.1012&-0.8394& 0.1910&&-0.6738& 0.1121&-0.8385& 0.1775&&-0.4138& 0.0842\\
\textbf{C$_7$}          &-0.1822&-0.1488&-0.2642&-0.1041&&-0.1749&-0.1610&-0.2538&-0.1186&&-0.1939&-0.1287&-0.2820&-0.0825&&-0.0800&-0.0554\\
\textbf{H$_7$}          & 0.1120& 0.0079& 0.1229& 0.0017&& 0.1172& 0.0121& 0.1293& 0.0058&& 0.1201& 0.0069& 0.1346&-0.0004&& 0.1162&-0.0119\\
\textbf{C$_8$}          &-0.2449& 0.0802&-0.1959& 0.0312&&-0.2677& 0.0925&-0.2179& 0.0379&&-0.2530& 0.0629&-0.1973&-0.0068&&-0.1193& 0.0060\\
\textbf{H$_8$}          & 0.1433& 0.0033& 0.1664& 0.0076&& 0.1518& 0.0027& 0.1760&-0.0076&& 0.1488& 0.0091& 0.1731& 0.0171&& 0.1064& 0.0031\\
\textbf{C$_\mathrm{8A}$}& 0.1389&-0.1907& 0.1075&-0.0992&& 0.1508&-0.2006& 0.1193&-0.1055&& 0.1382&-0.1472& 0.0967&-0.0218&& 0.1216& -0.0271\\
\hline\hline
\end{tabular}
  \label{TAB-cha} 
\end{table*}
Partial charges  using TD-DFT with different functionals and the aug-cc-pVTZ basis set  were calculated for the ground- and excited state (vertical electronic excitation) of 1MQ. 
The excited state of interest was chosen to be the first bright state which was $\mathbb{S}1$ throughout all calculations. 
The calculated partial charges of the ground state $\mathbb{S}0$ and their change $\Delta q_{j\gamma}$ upon excitation to $\mathbb{S}1$ are shown in Table~\ref{TAB-cha}. 
Upon excitation, electron density is shifted from the phenyl ring and the oxygen atom of 1MQ to the pyridinium part of the molecule, resulting in a lowering of the dipole moment by 30-40\%.
This finding was already reported by Sebastiani \textit{et al.}~\cite{seb11a}.
Table~\ref{TAB-dip} lists the obtained excitation wavelengths. 
With increasing complexity of the applied functional, as well as upon inclusion of implicit solvation by a polarizable continuum model (PCM), the computed excitation wavelength $\lambda_\mathrm{abs}$ resemble 
the experimental value of \SI{432}{\nano\meter} of 1MQ in water~\cite{ern05a,kov07a} more closely. Ref.~\cite{seb11a} also reported on the success of the PBE0 functional for the computation of excitation energies and referred to
a study~\cite{ada09a} comparing various organic dyes.

Although the individual charge distributions vary, the strength of the dipole moment $\mu$ is only slightly influenced by the use of different functionals, 
as can be seen in Table ~\ref{TAB-dip}. Dipole moments in vacuum were calculated to be 10-\SI{11}{D} in $\mathbb {S}0$ and decrease to about \SI{7}{D} upon excitation, which is in good
agreement to values reported in Ref.~\cite{seb11a}. Including PCM results in an increased dipole moment in both ground- and excited state, but leaves the ratio between them unchanged.
The orientation of the dipole moment is independent from the applied method and functional and does hardly change upon excitation.
Fig.~\ref{FIG:dipole} shows exemplarily the dipole moment of the ground state of 1MQ calculated using CHelpG B3LYP in vacuum. Plots of the dipole moment for all other methods and functionals 
for $\mathbb{S}0$ and $\mathbb{S}1$ look similar, apart from the differing strengths of $\mu$ (not shown).
\begin{table}[h!]
 \caption{\textcolor{black}{Quantum-mechanical results using the B3LYP, PBE0 and $\omega$B97xD functional with and without implicit solvation by a polarizable continuum model (PCM):     
 Computed excitation wavelengths $\lambda_\mathrm{abs}$ and 
 dipole moments $|\mu_j|$ and their components along the x, y and z axis (as shown in Fig.~\ref{FIG:dipole}) as well as the molecular polarizability $\alpha$.}}
 \centering
 \begin{tabular}{rrL{1.3cm}C{1cm}R{0.5cm}R{0.8cm}R{0.8cm}R{0.8cm}R{0.8cm}R{0.8cm}}
 \hline\hline
						&&&  $\lambda_\mathrm{abs}$		  && $\mu_x$&$\mu_y$& $\mu_z$& $|\mu|$ & $\alpha$ \\
						&& &[nm]		 && [D]    & [D]	  & [D] & [D] & [\AA$^3$] \\ 
						\hline 
\parbox[t]{2mm}{\multirow{4}{*}{\rotatebox[origin=c]{90}{\footnotesize CHelpG}}}& \parbox[t]{2mm}{\multirow{4}{*}{\rotatebox[origin=c]{90}{\footnotesize \textbf{B3LYP}}}}
		& \multirow{2}{*}{ vacuum}	&  \multirow{2}{*}{606}&$\mathbb{S}0$ 	&5.7			&8.9			&0.0			&\textbf{10.6} & \multirow{4}{*}{22.4}\\
					&&	& &$\mathbb{S}1$	&3.7			&6.3			&0.0			&\textbf{7.3}\\
&		& \multirow{2}{*}{ PCM}	& \multirow{2}{*}{522} &$\mathbb{S}0$	&8.0			&13.7			&0.0			&\textbf{15.9}\\
  					&&	& &$\mathbb{S}1$	&5.2			&8.4			&0.0			&\textbf{9.9}\\\\
\parbox[t]{2mm}{\multirow{4}{*}{\rotatebox[origin=c]{90}{\footnotesize CHelpG}}}& \parbox[t]{2mm}{\multirow{4}{*}{\rotatebox[origin=c]{90}{\footnotesize \textbf{PBE0}}}}
		& \multirow{2}{*}{ vacuum} & \multirow{2}{*}{586} &$\mathbb{S}0$	&5.7			&9.0			&0.0			&\textbf{10.6} &  \multirow{4}{*}{22.2}\\
  					&&	& &$\mathbb{S}1$	&3.7			&6.3			&0.0			&\textbf{7.3}\\
&		& \multirow{2}{*}{ PCM}	& \multirow{2}{*}{506} &$\mathbb{S}0$	&8.1			&14.3			&0.0			&\textbf{16.5}\\
  					&&	& &$\mathbb{S}1$	&5.2			&8.3			&0.0			&\textbf{9.8}\\\\
\parbox[t]{2mm}{\multirow{4}{*}{\rotatebox[origin=c]{90}{\footnotesize CHelpG}}}& \parbox[t]{2mm}{\multirow{4}{*}{\rotatebox[origin=c]{90}{\footnotesize \textbf{$\omega$B97xD}}}}
		& \multirow{2}{*}{ vacuum} &\multirow{2}{*}{541} &$\mathbb{S}0$ 	&5.9			&9.4			&0.0			&\textbf{11.1} & \multirow{4}{*}{22.1}\\
  					&&	& &$\mathbb{S}1$	&3.9			&5.9			&0.0			&\textbf{7.1}\\
&		& \multirow{2}{*}{ PCM}& \multirow{2}{*}{467} &$\mathbb{S}0$	&8.1			&14.3			&0.0			&\textbf{16.5}\\
  					&&	&&$\mathbb{S}1$	&5.5			&8.2			&0.0			&\textbf{9.9}\\
 \parbox[t]{2mm}{\multirow{4}{*}{\rotatebox[origin=c]{90}{\footnotesize Hirshf.}}}& \parbox[t]{2mm}{\multirow{4}{*}{\rotatebox[origin=c]{90}{\footnotesize \textbf{PBE0}}}}\\
&		& \multirow{2}{*}{ vacuum}	&&$\mathbb{S}0$ 	&5.7			&8.8			&0.0			&\textbf{10.5}\\
  					&&	&&$\mathbb{S}1$	&3.6			&6.1			&0.0			&\textbf{7.1}\\  
  \\\hline\hline
 \end{tabular}
 \label{TAB-dip} 
\end{table}
According to a simple dielectric model of a dipole embedded in its own high-frequency dielectric constant $\epsilon(\infty)$, the dipole moment $\mu_\mathrm{solv}$
\begin{equation}
 \mu_\mathrm{solv} = \frac{\epsilon(\infty) +2}{3} \cdot \mu_\mathrm{vacuum}
 \label{EQU:solv}
\end{equation}
increases due to the reaction field $\vec E^{RF}$ \cite{bor78b}. Using the average Voronoi \cite{oka00a} volume of 1MQ in our trajectories, $V$=\SI{224.1}{\cubic\angstrom}, and molecular polarizabilities $\alpha$
of 1MQ in $\mathbb{S}0$ obtained from frequency calculation in GAUSSIAN09 and tabulated in Table~\ref{TAB-dip}, the high frequency limit of the dielectric constant, $\epsilon(\infty)$, can be evaluated via
\begin{equation}
\frac{4 \pi \alpha}{3 V}  = \frac{\epsilon(\infty) -1}{\epsilon(\infty) +2}
\end{equation}
yielding \textcolor{black}{roughly 3.1. As a result, an average solvated dipole $\mu_\mathrm{solv}$ of \SI{18}{D} is expected according to Eq.~\eqref{EQU:solv}} which is a little bit higher than the quantum mechanical values in Table~\ref{TAB-dip}.

\textcolor{black}{To check for a correct hydrogen-bonding behavior of the solute, we performed a hydrogen-bond analysis exemplarily for the 1MQ with partial charges from $\omega$B97xD functional with PCM and SWM4 water 
and found that in the groundstate on average 3.3 water molecules are hydrogen-bonded to the oxyquinolinium oxygen, which is in good agreement with the 3.6 molecules for the $\mathbb{S}0$ state found in Ref.~\cite{ern11a}. 
In the excited state, the hydrogen-bonds are weakened and become less important~\cite{ern11a}. 
The solute model using our calculated partial charge distribution is therefore capable of describing correct hydrogen-bonding even though no explicit water molecules were included in the DFT calculation and the partial charges
therefore correspond to a solute immersed in a dielectric continuum representing the solute.
}

\subsection{Stokes shifts} 
\begin{figure}[ht!]
 \centering
 \includegraphics [width=8.5cm]{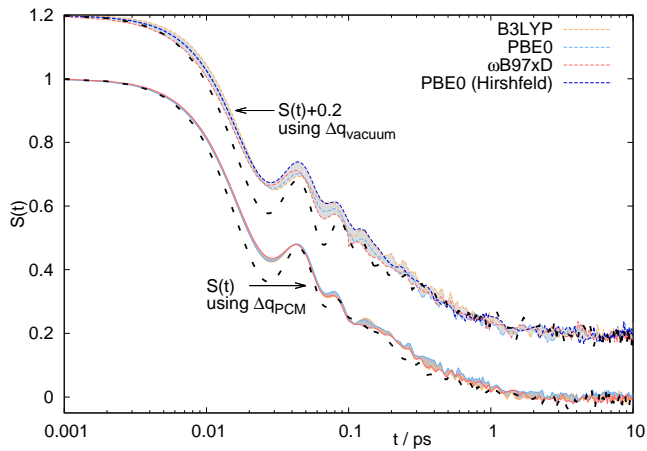}
 \caption{Stokes shift relaxation functions $S(t)$ of 1MQ in SPC water using the partial charge distributions from calculations in vacuum (dashed colored line) and in implicit solvent (solid colored line) obtained from $\Delta U$ via Eq.~\eqref{EQU:DeltaU}. 
           Note that for the sake of clarity the former is shifted by 0.2.
           The black dashed line represents the calculated Stokes shift from 1MQ in SPC water using the $\omega$B97xD partial charge distribution with and without PCM model obtained from $\Delta U$ via Eq.~\eqref{EQU:DeltaUq}.}
 \label{FIG:neq}
\end{figure}

Calculated Stokes shift relaxation functions $S(t)$ obtained from 1000 non-equilibrium simulations of 1MQ in SPC water per system using the various partial charge distributions in Table~\ref{TAB-cha} are shown in Fig.~\ref{FIG:neq}. 
Although the respective partial charge distributions differ, the solvation dynamics seem to be similar throughout all applied functionals and methods as visible by the small gray shaded areas in Fig.~\ref{FIG:neq}. Charges obtained from PCM models yielded
slightly faster relaxation than those from vacuum models, which might be an effect of the larger change in the dipole moment of the solute. 
Different functionals (B3LYP, PBE0 and $\omega$B97xD) have nearly no effect on the Stokes shift of 1MQ in water. 

In principle, $S(t)$ consists of processes on at least two different time scales~\cite{ern11a,mar07a}. The initial fast solvent response  can be represented by a Gaussian function and
the ``long-term'' relaxation by a Kohlrausch-Williams-Watt (KWW) function, so that the overall relaxation is modeled by
\begin{equation}
 S(t) \approx a e^{-(t/\tau_1)^2} + (1-a) e^{-(t/\tau_2)^\beta}
 \label{EQU:fit}
\end{equation}
yielding average relaxation times
\begin{equation}
 \langle\tau\rangle = a \frac{\tau_1}{2} \sqrt{\pi} + (1-a) \frac{\tau_2}{\beta} \Gamma \Bigl(\frac{1}{\beta} \Bigl)
\end{equation}
listed in Table~\ref{TAB-relax}.
\textcolor{black}{Experimental data is usually fitted by either multiexponential decay~\cite{ern11a,cas95a,voe03a,bha04a,kov07a} or stretched exponentials~\cite{mar15a,mar07d,mar03a,mar07a} and a Gaussian function only if
the solvent relaxes slow enough for the inertial part to be measured.
As the use of a stretched exponential instead of multiple exponentials reduces the number of fitting parameters, we decided to use the KWW function, as also published in Ref.~\cite{reg14a}.
}
As already indicated by Fig.~\ref{FIG:neq}, solutes with partial charge distribution calculated including PCM show slightly faster relaxation times than those  
calculated in vacuum. Averaged over all systems we obtained a relaxation time of \SI{0.15}{ps}, so that all simulations in non-polarizable SPC water relax too fast in comparison with experiment, 
where the average relaxation time was measured to be \SI{0.48}{\pico\second} by Sajadi \textit{et al.}~\cite{ern11a} at \SI{294}{K} and \SI{0.42}{\pico\second} at \SI{298}{K} 
by Perez \textit{et al.}~\cite{kov07a}. From data by Sebastiani and coworkers~\cite{seb13a} at \SI{303}{K} of 1MQ in D$_2$O, who found their data indistinguishable from H$_2$O data, we calculate a
relaxation time of \SI{0.28}{\pico\second}, which is still slower than our simulations in SPC or TIP4P water. 
Including polarizability in our simulation yields a relaxation time closer to experiment, namely \SI{0.24}{\pico\second}, as is listed in Table~\ref{TAB-relax} for SWM4 water. 
\begin{table*}[b]
 \caption{Kohlrausch and gaussian fit parameter, as indicated in Eq.~\eqref{EQU:fit}, and average relaxation times of the Stokes shift $S(t)$ of 1MQ in SWM4/SPC water \textcolor{black}{and total magnitudes
 $\Delta \Delta U$ of the Stokes shift obtained from the simulation or from Eq.~\eqref{EQU:OLM} (indexed OLM).}}
 \centering
 \begin{tabular}{lll R{1cm}R{1cm}R{1cm}R{1cm}R{1cm}R{1.5cm}R{1.2cm}R{2cm}R{2cm}}
 \hline\hline
 \multicolumn{3}{c}{\footnotesize Partial charge distribution}
& 	a	& $\tau_1$	& $\tau_2$& $\beta$	 & $\langle\tau\rangle$ 
                &$\Delta U(0)$ & $\Delta U(\infty)$ & $\Delta\Delta U$ & $\Delta\Delta U_\mathrm{OLM}$\\
 & & &               & [ps] & [ps] & & [ps] & [eV] & [eV] & [cm$^{-1}$] & [cm$^{-1}$]\\\hline\\
\multicolumn{11}{c}{non-polarizable \textbf{SPC} water } \\ 
\multicolumn{7}{l}{\underline{via $\Delta U$ from Eq. \eqref{EQU:DeltaU} (charge-charge interaction):}}\\[0.2cm]
 CHelpG & \textbf{B3LYP} &  vacuum	&0.42		&0.015	& 0.26		&0.80		&\textbf{0.17}	&0.44	&0.28	&1349 & \textcolor{black}{1001}\\
 CHelpG & \textbf{B3LYP} &  PCM		&0.40		&0.015	&0.17		&0.71		&\textbf{0.13}	&1.11	&0.67	&3527 & \textcolor{black}{3310}\\
 CHelpG & \textbf{PBE0} &  vacuum	&0.41		&0.014	&0.21 		&0.83		&\textbf{0.14}	&0.45	&0.27	&1425 & \textcolor{black}{1001}\\
 CHelpG & \textbf{PBE0} &  PCM		&0.40		&0.015	&0.18 		&0.67		&\textbf{0.15}	&1.28	&0.74	&4343 & \textcolor{black}{4127}\\
 Hirshfeld& \textbf{PBE0} &  vacuum	&0.38		&0.013	&0.22 		&0.83		&\textbf{0.16}	&0.36	&0.19	&1348 & \textcolor{black}{1063}\\
 CHelpG & \textbf{$\omega$B97xD}&vacuum	&0.36		&0.014	&0.18 		&0.71		&\textbf{0.15}	&0.54	&0.31	&1854 & \textcolor{black}{1471}\\
 CHelpG & \textbf{$\omega$B97xD} &  PCM	&0.39		&0.015	&0.16 		&0.69		&\textbf{0.13}	&1.24	&0.72	&4190 & \textcolor{black}{4005}\\
 \\ 
 \multicolumn{7}{l}{\underline{via $\Delta U$ from Eq. \eqref{EQU:DeltaUq} (dipole-charge interaction):}}\\[0.2cm]
 CHelpG & \textbf{$\omega$B97xD}&vacuum	&0.49		&0.014	&0.24 		&0.77		&\textbf{0.15}	&0.33	&0.17	&1343\\
 CHelpG & \textbf{$\omega$B97xD} &  PCM	&0.49		&0.013	&0.18 		&0.83		&\textbf{0.11}	&0.85	&0.43	&3394\\ 
 \\
 \multicolumn{11}{c}{\textcolor{black}{non-polarizable \textbf{TIP4P} water} }  \\
 \multicolumn{7}{l}{\textcolor{black}{\underline{via $\Delta U$ from Eq. \eqref{EQU:DeltaU} (charge-charge interaction):}}}\\[0.2cm]
 \textcolor{black}{CHelpG} & \textcolor{black}{\textbf{$\omega$B97xD}} &  \textcolor{black}{PCM}	&\textcolor{black}{0.37}&\textcolor{black}{0.015}&\textcolor{black}{0.18}&\textcolor{black}{0.65}&\textcolor{black}{\textbf{0.16}}&\textcolor{black}{1.26}&\textcolor{black}{0.74}&\textcolor{black}{4178} & \textcolor{black}{3980}\\
 \\
  \multicolumn{11}{c}{\textcolor{black}{non-polarizable \textbf{TIP4P/2005} water} } \\ 
  \multicolumn{7}{l}{\textcolor{black}{\underline{via $\Delta U$ from Eq. \eqref{EQU:DeltaU} (charge-charge interaction):}}}\\[0.2cm]
 \textcolor{black}{CHelpG} & \textcolor{black}{\textbf{$\omega$B97xD}} &  \textcolor{black}{PCM}	&\textcolor{black}{0.36}&\textcolor{black}{0.014}&\textcolor{black}{0.25}&\textcolor{black}{0.56}&\textcolor{black}{\textbf{0.27}}&\textcolor{black}{1.28}&\textcolor{black}{0.76}&\textcolor{black}{4262} & \textcolor{black}{3996}\\
 \\
 \multicolumn{11}{c}{polarizable \textbf{SWM4} water} & \\ \multicolumn{7}{l}{\underline{via $\Delta U$ from Eq. \eqref{EQU:DeltaU} (charge-charge interaction):}}\\[0.2cm]
 CHelpG & \textbf{$\omega$B97xD} &  PCM	&0.22		&0.021	&0.10  		&0.41		&\textbf{0.24}	&1.29	&0.74	&4453 & \textcolor{black}{3158}\footnotemark[4]\\
 \\
 \multicolumn{7}{l}{\underline{via $\Delta U$ from Eq. \eqref{EQU:DeltaUq} (dipole-charge interaction):}}\\[0.2cm]
 CHelpG & \textbf{$\omega$B97xD} &  PCM	&0.20		&0.016	&0.09  		&0.41		&\textbf{0.21}	&0.87	&0.43	&3539\\
 \\
 \\
 \multicolumn{11}{c}{ab initio MD simulation~\cite{seb13a}} \\
  & \textbf{PBE0}\footnotemark[1] & PCM &  0.29        & 0.052 & 0.49          & 0.95          & \textbf{0.37}  &       &       & 4033\\\\
 \multicolumn{11}{c}{experiment~\cite{ern11a,ern14a,sac00a,hes02a} }  & \\
       & & & 0.00 &  & 0.27 & 0.91 &  \textbf{0.28}\footnotemark[2] & & & $\sim$2500  \\ 
       & & & 0.00 &  & 0.35& 0.63 &  \textbf{0.48}\footnotemark[3] & & & 2300\\
  \hline\hline
 \end{tabular}
 \footnotetext[1]{triple-$\zeta$ Pople 6-311G**}
 \footnotetext[2]{\textcolor{black}{fit from experimental data at \SI{303}{K} as printed in Ref.~\cite{seb13a}, originally published in Ref.~\cite{ern11a}}}
 \footnotetext[3]{fit from data of Ref.~\cite{ern14a} at \SI{294}{K}}
 \footnotetext[4]{\textcolor{black}{In contrast to all other water models with $\epsilon(\infty)=1$, the polarizable SWM4 has a $\epsilon(\infty)=$1.4 reducing $\Delta\Delta_\mathrm{OLM}$ in Eq.~\eqref{EQU:OLM} from \SI{4020}{\per\centi\meter}
 to \SI{3158}{\per\centi\meter}.}}
 \label{TAB-relax} 
\end{table*}

Table~\ref{TAB-relax} also lists the difference in solvation energy after the excitation, $\Delta U(t=0)$, and after relaxation, $\Delta U(t=\infty)$, as well as the magnitude of the 
observed fluorescence shift $\Delta \Delta U$.
We found that systems using the PCM partial charges show a large shift of about \SI{4100}{\centi\meter^{-1}} which is in good agreement with the ab initio simulation using density functional theory 
in Ref.~\cite{seb11a,seb13a}. Systems using the vacuum partial charges show smaller shifts of about \SI{1500}{\centi\meter^{-1}}. 
\textcolor{black}{Based on the Ooshika-Lippert-Mataga theory (OLM) \cite{fle83a} the fluorescence shift is}
\begin{equation}
 \Delta \Delta U_\mathrm{OLM} = \frac{2 (\mu_{\mathbb{S}0}-\mu_{\mathbb{S}1})^2}{3 \epsilon_0 \cdot h c \cdot V} \Biggl( \frac{ \epsilon(0)-1}{2 \epsilon(0)+1} - \frac{ \epsilon(\infty)-1}{2 \epsilon(\infty)+1} \Biggl) 
 \label{EQU:OLM}
\end{equation}
\textcolor{black}{using the dielectric permittivity of the vacuum $\epsilon_0$=\SI{8.85E-12}{\ampere \second / \volt \meter} and the dielectric constant of the solvent at zero frequency $\epsilon(0)$ and at the 
high frequency limit $\epsilon(\infty)$ as well as the solute volume $V$. The dependence of $\Delta \Delta U$ on $\Delta \mu^2$ was also reported in Ref.~\cite{mar97a}. 
The predicted shift $\Delta \Delta U_\mathrm{OLM}$ in the non-polarizable water models fits quite well for all partial charge distributions obtained by using the polarizable continuum model as visible in Table~\ref{TAB-relax}.
Larger discrepancies are detected for the partial charge distributions calculated for 1MQ in vacuum and for $\Delta \Delta U_\mathrm{OLM}$ in the polarizable SWM4 model using the dielectric data from Ref.~\cite{sch15a}.
However, all these shifts are larger than the experimental observed Stokes shifts of 2300-\SI{2750}{\centi\meter^{-1}} in Ref.~\cite{kov07a,ern11a,seb13a,ern14a} showing that the latter do not account for the full shift
since these shifts started at the first time step that could be measured instead of $t=$\SI{0}{\pico\second}.}

\bigskip
The Stokes shift relaxation function calculated via the dipole approximation from Eq.~\eqref{EQU:DeltaUq} is also shown in Fig.~\ref{FIG:neq} (black dashed lines) for the partial charge distributions obtained via CHelpG $\omega$B97xD with and 
without PCM. The relaxation behavior is quite similar to the atomistic $S(t)$, however, the magnitude of the effect decreases to about three quarters of the original shift, which can be seen by the different values for $\Delta\Delta U$
in Table~\ref{TAB-relax}. The interaction energy of the initial non-equilibrium conformation of the system, as well as of the equilibrated conformation is smaller than the charge-charge Coulomb interaction energy. 
Nevertheless, the approximation of the solute as a dipole in a cave for the energy calculation holds true. 
\textcolor{black}{
Consequently, the solvation interaction is rather unaffected by the local charge density
of the solute atoms as long as the transition dipole moment from ground to excited state is represented reasonably.
It should be kept in mind though, that the approximation affects only the 
merging of the partial charge distribution into a single solute dipole moment used to calculate the Stokes shift, not the partial charge distribution used for 
the trajectory itself, which was obtained by simulating the atomistic solute. 
}

\subsection{Shell resolved Stokes shift}
\begin{figure}[ht]
 \includegraphics[width=8.5cm]{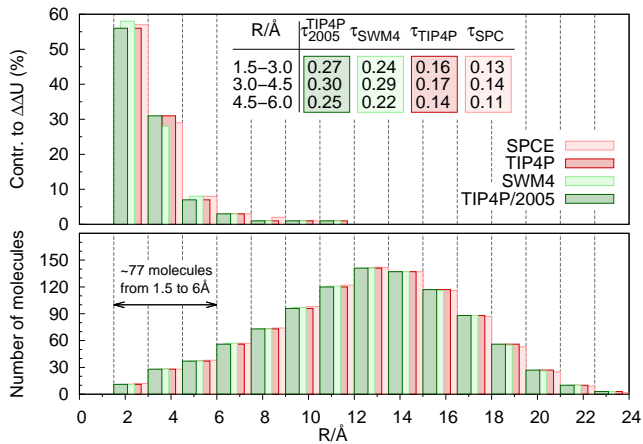}
 \caption{\textcolor{black}{Top: Contribution of the solvent atoms with their center of mass at the respective distance $R$ to the closest oxyquinolinium atom to the overall magnitude $\Delta \Delta U$. The inset lists an estimate of the 
 mean relaxation times $\langle \tau \rangle$ of the water molecules in the three bins that contribute most to the shift. Bottom: Number of
 solvent molecules at the respective distances.}}
 \label{FIG:perc}
\end{figure}

To gain more insights into solvent properties, we decomposed the Stokes shift to its contributions from the respective shells around the solute molecule as shown in Table~\ref{TAB-shell}.
About 85\% of the Stokes shift comes from solvent molecules in the first hydration shell of 1MQ, 12\% from the second shell and 2\% from the third shell. 
The contribution from more remote shells is negligible. 
Although 97\% of the magnitude of the observable Stokes shift stems from the first two solvation shells, the simulated system should also contain the 
remote shell to some extent in order to avoid computational artifacts from the periodic boundary conditions.
\begin{table}[ht]
 \centering
 \caption{Contributions of the respective solvent shells around 1MQ to the Stokes shift calculated using the partial charges from the CHELPG $\omega$B97xD PCM method.}
 \begin{tabular}{l R{1.5cm}R{1.5cm}R{1.5cm}R{1.5cm}}
 \hline\hline
	    & shell 1& shell 2& shell 3& rest \\\hline
 SPC	    & 83.8\% & 12.9\% & 2.4\% &0.8\%\\
 TIP4P      & 85.4\%  & 11.7\%  & 2.3\%  & 0.6\% \\
 TIP4P/2005 & 88.3\% & 8.9\% &  2.2\% &  0.6\% \\
 SWM4	    & 86.3\% & 11.3\% & 2.3\% &0.2\%\\
  \hline\hline\
 \end{tabular}
 \label{TAB-shell} 
\end{table}
\textcolor{black}{We furthermore analyzed the contributions of solvent molecules at different distances from the closest chromophore atom in a more detailed analysis and found that about 58\% 
from the shift comes from SWM4 water at distances 1.5 to \SI{3}{\angstrom} with a relaxation time of roughly \SI{0.24}{ps},
29\%  at distances 3 to \SI{4.5}{\angstrom} with $\langle \tau \rangle$ of about \SI{0.29}{ps}, 8\%  at distances 4.5 to \SI{6}{\angstrom} with a relaxation time of \SI{0.22}{ps} and 8\% 
 from water molecules further away as shown in Fig.~\ref{FIG:perc}. For SPC, TIP4P and TIP4P/2005 water the contributions are nearly the same, indicating no structural changes, but with relaxation times of 0.11 to \SI{0.14}{ps} for SPC, 
0.14 to \SI{0.17}{ps} for TIP4P and 0.25 to \SI{0.30}{ps} for TIP4P/2005,
indicating differences in dynamical properties. Although the decomposition
of the overall shift into its contributions lowers the statistics and the calculated relaxation times are only rough estimates, still a rather uniform relaxation behavior independent of the distance to the chromophore can be observed, so
that the oxyquinolinium serves as a good probe of bulk water properties, as it does nearly not change the relaxation time of the solvent molecules close to it.}
\textcolor{black}{Fig.~\ref{FIG:perc} also shows the number of water molecules contributing to each bin. The three histogram bins with the highest contributions
to the magnitude of the Stokes shift contain about 77 water molecules, which corresponds to the complete first shell and parts of the second shell.}

\subsection{Influence of the solvent}
\begin{figure}[b]
 \centering
 \includegraphics [width=8.5cm]{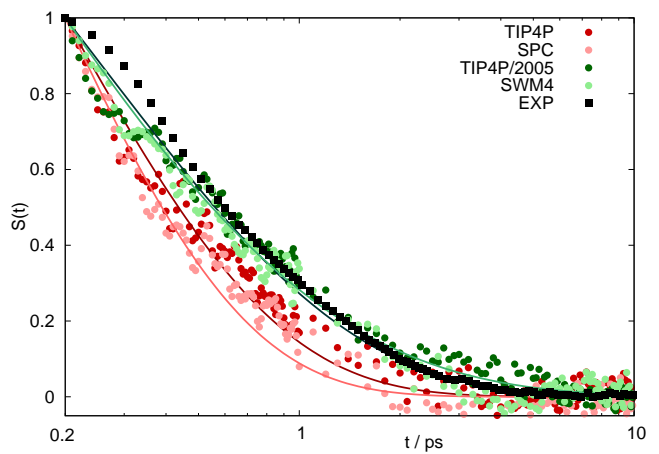}
 \caption{Comparison of experimental (EXP) data, extracted from Ref.~\cite{ern14a}, with the Stokes shift of 1MQ in SPC and SWM4 water using the CHelpG $\omega$B97xD 
 partial charge distribution 
 from calculations in implicit solvent. Solid lines represent the 
 respective fitted Kohlrausch/Gaussian functions (see Eq.~\eqref{EQU:fit}).}
 \label{FIG:neq-swm4}
\end{figure}

Fig.~\ref{FIG:neq-swm4} shows the calculated Stokes shift of 1MQ in SPC, \textcolor{black}{TIP4P, TIP4P/2005} and SWM4 water respectively, using the partial charge distribution CHelpG $\omega$B97xD PCM as 
well as experimental data from Ernsting and coworkers~\cite{ern14a}. To ensure comparability, the computed Stokes shift was set to 1 at t=\SI{0.2}{\pico\second}. 
The solid lines represent the fit according to Eq.~\eqref{EQU:fit} with the parameters given in Table~\ref{TAB-relax}.
The Stokes shift obtained from simulation in polarizable SWM4 water comes closer to experiment than in the faster non-polarizable SPC water, but is still slightly too fast. 
However, only the initial response shows small differences between our simulation and experiment, the long-term relaxation agrees quite well. 
\textcolor{black}{Analogue to SPC, the TIP4P water model shows too fast dynamics. The newer TIP4P/2005 model in contrast yields relaxation times similar to the polarizable SWM4 water.}

\bigskip
The average relaxation time depends on the solvent viscosity $\eta$ which itself depends on the diffusion coefficient $D$ and the rotational relaxation constant 
$\tau_{\mathrm{rot}}$~\cite{ste14c} via
\begin{equation}
 \displaystyle \eta \simeq \frac{k_B T}{3 \pi} \cdot \frac{1}{\sqrt{6 D^3 \tau_{\mathrm{rot}}}}.
 \label{EQU-visc}
\end{equation}
\textcolor{black}{To investigate whether the different relaxation behavior of the four water models stems from different solvent viscosity, we calculated the $D$ and $\tau_{\mathrm{rot}}$ for SPC and SWM4 water, respectively.}
The diffusion coefficient computed from the mean-square-displacement of an equilibrium simulation is \SI{0.25}{\square\angstrom\per\pico\second} 
for SWM4 water and \SI{0.43}{\square\angstrom\per\pico\second} for SPC water. 
The SWM4 model comes quite close to the experimental value which was determined to be \SI{0.23}{\square\angstrom\per\pico\second} by Sacco \textit{et al.}~\cite{sac00a}, but the SPC model shows too
fast dynamics. Calculation of $\langle {\mu}(0) \cdot {\mu(t)} \rangle$ of an equilibrium simulation and exponential fitting then yields the fit parameter
$\tau_{\mathrm{rot}}$, which is \SI{3.0}{\pico\second} for SPC water and \SI{4.7}{\pico\second} for SWM4 water, respectively. By inserting these values into Eq.~\eqref{EQU-visc} we get a viscosity
of \SI{0.4}{mPas} for SPC and \SI{0.7}{mPas} for SWM4 water (see Table~\ref{TAB-transport}) which coincide with previously published data for SPC  \cite{hes02a} and SWM4 water \cite{ker05c}. 
This finding is in agreement with J\"{o}nsson \textit{et al.} who reported that adding electronic polarizability to a water model slows down dynamic properties and increases the viscosity~\cite{joe89a}. 
Experiments measured the viscosity at \SI{300}{\kelvin} to be \SI{0.85}{mPas}~\cite{hes02a}, so that \textcolor{black}{all} employed water models (\textcolor{black}{except for TIP4P/2005}) show too low viscosity and 
too large diffusion constants.
However, $\langle \tau \rangle$ in our computational water models and experimental data correlates roughly with the respective viscosities $\eta$ as already found quite generally in Ref.~\cite{mar03a}, 
so that the faster relaxation times in Table~\ref{TAB-relax} of our simulations compared to experiment are (at least partly) due to underestimated viscosities.

\begin{table}
 \caption{\textcolor{black}{Transport and dielectric properties \cite{sch15a} of water. }}
 \centering
 \begin{tabular}{lccccccc}
 \hline\hline
  & $D$ & $\tau_\mathrm{rot}$ & $\eta$ & $\epsilon(0)$ & $\epsilon(\infty)$ &$\tau_D$\footnotemark[1]& $\langle\tau\rangle_\mathrm{DS}$ \\ 
  & [\AA$^2$/ps] & [ps] & [mPa s] &             &                    &[ps] & [ps]\\\hline 
  SPC        & 0.43 & 3.0 & 0.4 & 65.7 & 1 & 6.6 & 0.15\\
  TIP4P      & 0.32\footnotemark[2] &     &  0.50\footnotemark[3] & 51.4 & 1 & 6.6 & 0.19\\
  TIP4P/2005 & 0.21\footnotemark[2] &     &  0.86\footnotemark[3] & 59.6 & 1 & 11.7 & 0.29\\    
  SWM4       & 0.25 & 4.7 & 0.7  & 78.9 & 1.4 & 10.9 & 0.26\\
  experiment & 0.23&     & 0.85 & 78.4 & 1.8 & 8.3 & 0.24\\
  \hline\hline
 \end{tabular}
 \footnotetext[1]{\textcolor{black}{$\tau_D=\tau$ from the Cole-Cole fit in Ref.~\cite{sch15a} since the $\gamma$'s are close to unity and the reciprocal value of both relaxation times determines
 the frequency of the respective dielectric peak maximum \cite{bor78a}.}}
 \footnotetext[2]{\textcolor{black}{see Ref.~\cite{ara09a}.}}
 \footnotetext[3]{\textcolor{black}{see Ref.~\cite{aba10a}.}}
 \label{TAB-transport}
\end{table}
\textcolor{black}{In contrast to dielectric spectroscopy, which probes collective polarization of the solvent, the relaxation of the Stokes shift is a measure of the local polarization.
However, the average relaxation time of the Stokes shift $\langle \tau \rangle_\mathrm{DS}$ can also be estimated on the basis of the dielectric spectrum (DS) of the solvent via}
\begin{equation}
 \langle \tau \rangle_\mathrm{DS} \simeq \frac{\epsilon_c + 2 \epsilon(\infty)}{\epsilon_c + 2 \epsilon(0)} \cdot \tau_D
\end{equation}
\textcolor{black}{using the high-frequency dielectric constant $\epsilon_c$ of the solute in a cavity \cite{fle83a,ern13a}. Based on the computational dielectric constants $\epsilon(0)$
and $\epsilon(\infty)$ in Ref.~\cite{sch15a} and extrapolating the Cole-Cole fit in that reference to a Debye process with the relaxation constant $\tau_D$, one gets a Stokes 
relaxation time $\langle \tau \rangle_\mathrm{DS}$ of \SI{0.15}{\pico\second},\SI{0.19}{\pico\second},\SI{0.29}{\pico\second} and \SI{0.26}{\pico\second} for SPC, TIP4P, TIP4P/2005 and SWM4 (see Table~\ref{TAB-transport}), 
respectively, which agree well with the Stokes relaxation times in Table~\ref{TAB-relax}. 
It should be noted that although the Stokes shift and the calculated relaxation times are comparable for the SWM4 and the TIP4P/2005 water model and both models are therefore fit for calculating solvation dynamics, the latter one fails to describe
\textcolor{black}{some} dielectric processes, as all non-polarizable
solvent models are characterized by a $\epsilon(\infty)$ of 1, where experiments yield $\epsilon(\infty)$ of 1.8.
For all solvent models, $\epsilon_c$ was set to 1 since a non-polarizable 1MQ was used in the simulations. However, if $\epsilon_c$ is increased to 
$\epsilon(\infty)=$3.1 obtained from the polarizability $\alpha$ and the Voronoi volume in a previous section, the corresponding $\langle \tau \rangle$ raise to \SI{0.25}{\pico\second}, \SI{0.32}{\pico\second}, \SI{0.49}{\pico\second}
and \SI{0.40}{\pico\second} for SPC, TIP4P, TIP4P/2005 and SWM4. These values are closer to the experimental results and argue for a polarizable solute during the non-equilibrium MD simulations
to determine the Stokes shift. This will be the topic of a future publication.
}

\section{Conclusion}
In this paper we  examined the effect of the partial charge distribution in ground- and excited state of the chromophore 1-methyl-6-oxyquinolinium betaine in water on the solvation dynamics obtained
from non-equilibrium MD simulation. For partial charges using the B3LYP, PBE0 or $\omega$B97xD functional and the CHelpG or Hirshfeld method in vacuum we found that varying charge distributions showed the same dipole moment 
concerning strength and orientation and hardly any differences in the Stokes shift relaxation function $S(t)$. 
The same holds true for all charge distributions obtained in implicit solvent: The dipole moments obtained from partial charge calculation via B3LYP, PBE0 or $\omega$B97xD functional and the CHelpG method in implicit solvent 
do not vary with the functional and the respective $S(t)$ are almost identical. 
However, minor differences could be found between calculations in vacuum and implicit solvent. The dipole moment is larger when using the PCM model, as well as the observed magnitude of
the overall Stokes shift resulting in a slightly faster relaxation. 

\bigskip
\textcolor{black}{Furthermore we applied different water models to investigate the effect of polarizability and force field parametrization of the solvent on the Stokes shift.
The inclusion of polarizability in the SWM4 water model slowed down the solvation dynamics (compared to SPC and TIP4P water)
as expected by the increased viscosity $\eta$, so that longer relaxation times, which were closer to 
experimental data, could be observed. Analogue results were obtained when applying the TIP4P/2005 model,
indicating that for the relaxation of the Stokes shift computational transport properties close to experiment are most important. This can be realized by introducing polarizable
forces (SWM4) or by a re-parametrization of the partial charges and/or Lennard Jones parameters (TIP4P/2005),
\textcolor{black}{giving equally accurate results concerning solvation dynamics. If further dielectric properties, \textit{e.g.} dielectric spectra, need to be computed we nevertheless recommend
only the polarizable water model, as it reproduces the correct dielectric constant at the high frequency limit.}
}
Decomposition of the Stokes shift into its contributions from different solvation shells via Voronoi tessellation exhibits that the Stokes shift is almost restricted to
two solvation shells \textcolor{black}{for all water models}. 
When comparing our results to experiment, we find that the agreement is in general very high. However, the long term relaxation process is in slightly better agreement than the initial fast response. 
This may be an effect of the assumptions made on the excited state, namely that the geometry and force-field parameters are not allowed to change which may influence the vibrational relaxation of
the excited state.
Nevertheless, the agreement is good enough to allow for predictions of the relaxation time, as well as of the magnitude and shape of the Stokes shift. \textcolor{black}{In addition, we showed that
the average relaxation time $\langle \tau \rangle$ of the Stokes shift can also be extrapolated from the static and high-frequency limit of the dielectric constant as well as the Debye relaxation time
of the solvent.}

\bigskip
We also calculated the Stokes shift obtained via the dipole approximation for the interaction energy and found that it shows nearly the same relaxation time as the Coulomb interaction energy. The dipole 
approximation apparently holds true for the 1MQ-water system and yields results close to the atomistic description of the solute with the respective $\Delta \vec \mu_j$. 
However, the absolute energies are shifted to lower interaction energies, and the magnitude of the overall shift is decreased to roughly three quarters, so that using a
dipole in a cavity instead of the true atomistic solute molecule may not be the method of choice for calculating absolute Stokes shifts. However, the fact that the solute may be represented by a dipole
instead of an exact partial charge distribution points out that the Stokes shift in our system does not depend on peculiarities of the solute.

\bigskip
Overall, the influence of the solute and its partial charge distributions in ground and excited state is rather small compared to the impact of the solvent. This points out that the Stokes shift is more a measure 
of the solvation and transport properties of the solvent.

\section{Supplementary Material}
See supplementary material for the CHARMM force field parameters and the respective coordinates in PDB format of the chromophore 1MQ.

\section{Acknowledgement}
We thank N. P. Ernsting and O. Steinhauser for critical reading of the manuscript.
This work was funded by the Austrian Science Fund FWF in the context of Project 
No. FWF-P28556-N34.


\end{document}